\documentclass[twocolumn,aps,prl,superscriptaddress,unsortedaddress]{revtex4}
\usepackage{amsmath}
\usepackage{graphicx}
\usepackage{dcolumn}
\usepackage{bm}
\usepackage{color}

\begin{document}
\title{Comment on ``Dynamical mean field solution of the Bose-Hubbard model''}
\author{K.~Byczuk}
\affiliation{Institute~of~Theoretical~Physics,~University~of~Warsaw,~ul.~Ho\.za~69,~PL-00-681~Warszawa,~Poland}
\author{D.~Vollhardt}
\affiliation{Theoretical Physics III, Center for Electronic Correlations and Magnetism,
Institute for Physics, University of Augsburg, D-86135 Augsburg, Germany}

\begin{abstract}
In their preprint Anders \emph{et al.}  [arXiv:1004.0510] propose a
crucial modification of the Bosonic Dynamical Mean-Field Theory
(B-DMFT)  derived by us [Phys. Rev. B {\bf 77}, 235106 (2008)]. Here
we show that the modification consists of two steps which, in fact,
cancel each other. Consequently their self-consistency equations are
identical to ours.
 \end{abstract}

\maketitle

According to  Anders \emph{et al.} \cite{Anders10} they have
modified the Bosonic Dynamical Mean-Field Theory (B-DMFT)
\cite{Byczuk08} ``to avoid double counting of kinetic energies'' and
thereby found stable solutions to the B-DMFT equations.
Here we show that the alleged double counting is caused by an
improper definition by Anders \emph{et al.} \cite{Anders10} of the
condensate wave function. In Ref. \cite{Anders10} the error
introduced thereby is subsequently compensated by (mathematically
uncontrolled) Bogoliubov shifts of the integration variables.

For simplicity we only consider the non-interacting case ($U=0$)
with nearest-neighbor hopping of the bosons on a lattice with finite
coordination number $Z$ since this is sufficient to illustrate our
point. The generalization to finite $U$ is straightforward. The
B-DMFT maps the lattice problem onto a single site (say, $i=0$)
which is self-consistently coupled to two dynamical baths, one
composed of normal and the other of condensed bosons
\cite{Byczuk08}. For all coordination numbers $2\leq Z < \infty$ the
exact partition function may be derived from a path integral with
the local action $Z_{\rm exact}=Z^{(0)} \int D[b_0^* b_0]
\exp(-S_{0}[b_0^* b_0])$. Here a superscript $(0)$ refers to a
lattice with a cavity, i.e., where the impurity site $i=0$ is
removed, while a subscript $0$ refers to the impurity site itself.
The local action for $U=0$ reads
\begin{eqnarray}
S_0= \int d\tau  b_0^*(\tau) (\partial _{\tau} - \mu) b_0(\tau)  \nonumber \\
+ \left[ \kappa \phi \int d \tau_1 b_0^*(\tau_1)  + H.c.\right] \nonumber \\
- \int d \tau_1 \int d \tau_2 \Delta (\tau_1-\tau_2)  b_0^*(\tau_1)  b_0(\tau_2),
\label{action}
\end{eqnarray}
where the condensate wave function
\begin{eqnarray}
\phi = \langle b_j \rangle_{S^{(0)}}
\end{eqnarray}
and the hybridization
function
\begin{eqnarray}\Delta(\tau_1-\tau_2) = \sum_{j_1 j_2} t_{0j_1} t_{j_20} \langle
b_{j_1}(\tau_1) b^*_{j_2}(\tau_2)\rangle_{S^{(0)}}
\end{eqnarray}
are determined
\emph{on the lattice with the cavity}, and $\kappa = \sum_j
t_{0j}=Zt$ is a geometric factor; cf. Appendix A of Ref.
\cite{Byczuk08} for a derivation and notation. In particular we note
that expectation values $\langle \dots\rangle_{S^{(0)}}$ only
contain connected terms.

In Ref. \cite{Anders10} a different approach is taken. Here the
expectation value of the boson is calculated \emph{on the impurity
site $i=0$}, $\langle b_0 \rangle_{S_0}$, and this quantity is
incorrectly identified with the condensate wave function $\phi$. By
performing Bogoliubov shifts of the integration variables $b_0$ and
$b_{0}^*$ the corresponding error in $\phi$ is compensated, whereby
the geometric term $\kappa$ now acquires an additional, but spurious
term linear in the hybridization function, and thus takes the form
$\kappa = Zt - \int d \tau_2\Delta(\tau_1-\tau_2)$. For this reason
the authors of \cite{Anders10} refer to $\kappa$ as ``the coupling
between the impurity and the condensate''.

It should be noted, however, that on a Bravais lattice  with
$2<Z<\infty$ the two averages are \emph{not} the same, i.e.,
$\langle b_j \rangle_{S^{(0)}}\neq \langle b_0 \rangle_{S_0}$. This
fact was already noted in Ref.~\cite{Hubner09} and was employed by
them to solve the B-DMFT equations numerically. In fact, as proved
in the Appendix, the two local averages obey the following exact
relation in Matsubara representation:
\begin{equation}
\langle b_j \rangle_{S^{(0)}}= \langle b_0 \rangle_{S_0}\left(1-\frac{G_{j0}}{G_{00}} \right),
\label{theorem}
\end{equation}
where $G_{ij}$ are two-site correlation functions on the
corresponding translationally invariant lattice. Using the correct
definition of $\phi$, namely  $\phi = \langle b_j
\rangle_{S^{(0)}}$, and the identity $\sum_{j_1}t_{0j_1} G_{j_1
0}(\omega_n) = -1 + \omega_n G_{00}(\omega_n)$ \cite{Georges96} it
immediately follows that the spurious hybridization term in $\kappa$
introduced by the Bogoliubov shifts in Ref. \cite{Anders10}
disappears again, whereby their self-consistency equations become
identical to ours \cite{Byczuk08}. The claims by Anders \emph{et
al.} \cite{Anders10} concerning deficiencies of the original B-DMFT
\cite{Byczuk08} are therefore invalid. In fact, their numerical
results \cite{Anders10} are the solution of the B-DMFT equations
previously derived  by us in Ref. \cite{Byczuk08}.

We thank Philipp Werner for sending us the manuscript plus
unpublished notes by L.~Pollet before submission of their preprint.
This work was supported in part by the TRR80 of the Deutsche
Forschungsgemeinschaft. KB was also supported by the grant N N202
103138 of Polish Ministry of Science and Education.

\subsection{Appendix: Proof of eq. (\ref{theorem})}

The local correlation functions are determined via a functional
derivative with respect to infinitesimally weak $U(1)$ symmetry
breaking fields $\eta_i$ which are added to the action, i.e., $
\langle b_{j}(\tau) \rangle _{S^{(0)}} = \delta \ln Z^{(0)} / \delta
\eta_{j}^*$.
The second
derivative gives the two-site correlation functions
\begin{equation}
\langle b_{j_1}(\tau_1) b^*_{j_2}(\tau_2)\rangle_{S^{(0)}} =
\frac{\delta^2 \ln Z^{(0)}}{\delta \eta_{j_1}^* \delta \eta_{j_2}} =
\frac{\delta }{\delta \eta_{j_2}} \langle b_{j_1}(\tau_1) \rangle
_{S^{(0)}}. \label{differential}
\end{equation}
In the   linear regime of  $\eta_i$ the solution of
(\ref{differential}) reads $ \langle b_{j_1}(\tau_1) \rangle
_{S^{(0)}} = \sum_{j_2} \langle b_{j_1}(\tau_1)
b^*_{j_2}(\tau_2)\rangle_{S^{(0)}} \eta_{j_2}$. Employing the exact
relation  between two-site correlation functions defined on a
lattice with a cavity, $G_{j_1j_2}^{(0)}$ , and on a translationally
invariant lattice, $G_{j_1j_2}$, respectively, namely
$G_{j_1j_2}^{(0)}= G_{j_1j_2} - G_{j_10}G_{00}^{-1} G_{0j_2}$
\cite{Georges96},  we find
\begin{eqnarray}
\langle b_{j_1} \rangle _ {S^{(0)}} &=& \sum_{j_2}(G_{j_1j_2} \eta
_{j_2} - G_{j_10}G_{00}^{-1} G_{0j_2} \eta_{j_2}) \nonumber \\
&=& \phi_{j_1} - G_{j_10}G_{00}^{-1} \phi_0,
\end{eqnarray}
where  $\phi_{j_1}$ and $\phi_0$ are local correlation functions at
site $j_1$ and $0$, respectively, on a translationally invariant
 lattice. In view of the translational invariance  these quantities are the same, i.e.,
$\phi_{j_1}=\phi_0$, which proves Eq.~(\ref{theorem}).

\end{document}